\documentclass[sn-mathphys-num]{sn-jnl}

\usepackage{graphicx}%
\usepackage{multirow}%
\usepackage{amsmath,amssymb,amsfonts}%
\usepackage{amsthm}%
\usepackage{mathrsfs}%
\usepackage[title]{appendix}%
\usepackage{breqn}
\usepackage{xcolor}%
\usepackage{textcomp}%
\usepackage{manyfoot}%
\usepackage{booktabs}%
\usepackage{algorithm}%
\usepackage{algorithmicx}%
\usepackage{algpseudocode}%
\usepackage{listings}%

\usepackage[utf8]{inputenc}
\DeclareUnicodeCharacter{2320}{/}
\DeclareUnicodeCharacter{23AE}{|}
\DeclareUnicodeCharacter{2321}{/}
\DeclareUnicodeCharacter{2572}{\textbackslash}
\DeclareUnicodeCharacter{2571}{/}
\DeclareUnicodeCharacter{3C6}{P}
\DeclareUnicodeCharacter{2080}{0}
\DeclareUnicodeCharacter{2208}{$\in$}
\DeclareUnicodeCharacter{22A5}{$\perp$}
\DeclareUnicodeCharacter{2220}{$\angle$}
\DeclareUnicodeCharacter{2225}{$\parallel$}
\DeclareUnicodeCharacter{225F}{$\stackrel{?}{=}$}
\usepackage{fancyvrb}
\usepackage{fvextra}
\usepackage[most]{tcolorbox}

\usepackage{pgfplots}
\pgfplotsset{compat=1.15}
\usepackage{mathrsfs}
\usetikzlibrary{arrows}

\usepackage{listings}

\definecolor{mygreen}{rgb}{0,0.2,0}
\definecolor{mygray}{rgb}{0.95,0.95,0.95}
\definecolor{mymauve}{rgb}{0.58,0,0.82}

\definecolor{lbcolor}{rgb}{0.95,0.95,0.95}

\lstdefinelanguage{giac}{
  keywords={factor, eliminate, solve, coeff, degree, csolve, list, sqrt},
  ndkeywords={>>},
  sensitive=true
}

\lstdefinelanguage{maple}{
  keywords={with, EliminationIdeal, PolynomialIdeals},
  sensitive=true
}

\theoremstyle{thmstyleone}%
\theoremstyle{thmstyletwo}%

\theoremstyle{thmstylethree}%

\begin{document}

\title[Automated proving based on the complex number identity method]
{Automated proving in planar geometry based on the complex number identity method and elimination}


\author*[1]{\fnm{Zolt\'an} \sur{Kov\'acs}}\email{zoltan@geogebra.org}
\equalcont{These authors contributed equally to this work.}

\author[2]{\fnm{Xicheng}\sur{Peng}}\email{pxc417@126.com}
\equalcont{These authors contributed equally to this work.}

\affil[1]{\orgname{Private University of Education, Diocese Linz}, \orgaddress{\street{Salesianumweg 3}, \city{Linz}, \postcode{4020}, \country{Austria}}}

\affil[2]{\orgdiv{National Engineering Laboratory for Educational Big Data},
\orgname{Central China Normal University}, \orgaddress{\street{No.~152, Luo Yu Road, Hongshan District},
\city{Wuhan}, \postcode{430079}, \country{China}}}


\abstract{
We improve the complex number identity proving method to a fully automated procedure, based on elimination
ideals. By using declarative equations or rewriting each real-relational hypothesis $h_i$ to $h_i-r_i$,
and the thesis $t$ to $t-r$, clearing the denominators and introducing an extra expression
with a slack variable, we eliminate all free and relational point variables. From the obtained
ideal $I$ in $\mathbb{Q}[r,r_1,r_2,\ldots]$ we can find a conclusive result. It plays an important role
that if $r_1,r_2,\ldots$ are real, $r$ must also be real if there is a linear polynomial $p(r)\in I$,
unless division by zero occurs when expressing $r$.
Our results are presented in Mathematica, Maple and
in a new version of the Giac computer algebra system. Finally, we present a prototype of the automated
procedure in an experimental version of the dynamic geometry software GeoGebra.
}




\maketitle

\section{Introduction}
\label{sec:introduction}

The complex number identity method \cite{PengZhangChenLiu_2023} is a well-known manual way to prove
problems in planar geometry, especially in the community of participating students and trainers
of the International Mathematical Olympiad and related contests \cite{ZhangPeng_2024}. It focuses
on free and constructed points in $\mathbb{C}$, and all dependent objects like angles or circles
are derived only by points and corresponding formulas.

In this paper we extend the method towards fully automatization, based on elimination theory from algebraic geometry
(see, for example, \cite{Cox_2007}, for a detailed introduction).
In fact, this extension was already started by the second author a couple of years ago, but it
was not yet fully elaborated. We fill some gaps in the underlying theory and present a more complete method
that can be mechanized. We illustrate the method in the free computer algebra system Giac and the well-known
Mathematica software (which is also available at no cost in the WolframScript framework),
and in Maple. Finally, we report on successful implementation of the method in GeoGebra Discovery \cite{ACM2021},
an experimental version of the free dynamic geometry software GeoGebra \cite{gg}.
Thus, our contribution is fully reproducible for a wide audience.

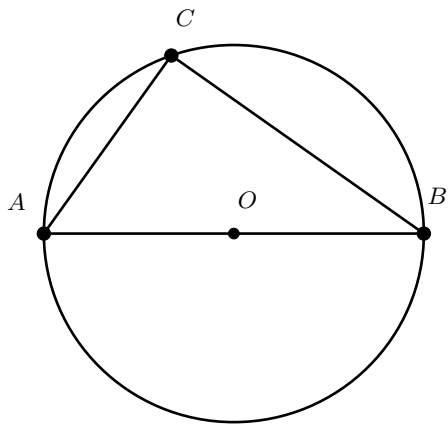
\begin{figure}[ht]
\definecolor{xdxdff}{rgb}{0,0,0}
\definecolor{uuuuuu}{rgb}{0,0,0}
\definecolor{ududff}{rgb}{0,0,0}
\begin{tikzpicture}[line cap=round,line join=round,>=triangle 45,x=5.0cm,y=5.0cm]
\clip(0.8084213835580526,0.31619736799461534) rectangle (2.2038497428722894,1.6790476333170805);
\draw [line width=1.pt] (1.,1.)-- (2.,1.);
\draw [line width=1.pt] (1.5,1.) circle (2.5cm);
\draw [line width=1.pt] (1.,1.)-- (1.3355337396563132,1.4721767139626463);
\draw [line width=1.pt] (2.,1.)-- (1.3355337396563132,1.4721767139626463);
\begin{scriptsize}
\draw [fill=ududff] (1.,1.) circle (2.5pt);
\draw[color=ududff] (0.9278743948612169,1.08449741796724) node {$A$};
\draw [fill=ududff] (2.,1.) circle (2.5pt);
\draw[color=ududff] (2.035529590581467,1.100786464963126) node {$B$};
\draw [fill=uuuuuu] (1.5,1.) circle (2.0pt);
\draw[color=uuuuuu] (1.5359988160409621,1.089927100299202) node {$O$};
\draw [fill=xdxdff] (1.3355337396563132,1.4721767139626463) circle (2.5pt);
\draw[color=xdxdff] (1.3731083460821019,1.5731688278438212) node {$C$};
\end{scriptsize}
\end{tikzpicture}
\caption{Examples 1 and 2: The converse of Thales' Circle Theorem}
\label{fig:12}
\end{figure}

We illustrate the method on the converse of Thales' Circle Theorem (Example 1).
Suppose $O$ is the midpoint of $AB$, and $C$ is a point on the circle with diameter $AB$ (Fig.~\ref{fig:12}).
By using items of the complex number identity method, we know that $A$, $O$ and $B$ are collinear, that is
\begin{align}\label{coll(A,O,B)}
\frac{A-O}{O-B}\in\mathbb{R},
\end{align}
and since $OA=OC$ the triangle $OAC$ is isosceles,
\begin{align}\label{equid(O,A,C)}
\frac{\frac{A-C}{A-O}}{\frac{C-O}{C-A}}\in\mathbb{R},
\end{align}
furthermore, since $OB=OC$ the triangle $OBC$ is also isosceles,
\begin{align}\label{equid(O,B,C)}
\frac{\frac{B-O}{B-C}}{\frac{C-B}{C-O}}\in\mathbb{R}.
\end{align}
We want to conclude that $AC\perp BC$, that is,
\begin{align}\label{perp(A,C,B,C)}
\left(\frac{C-B}{C-A}\right)^2\in\mathbb{R}.
\end{align}

In fact, the last expression (\ref{perp(A,C,B,C)}) (the thesis) means that either $AC\perp BC$ or $AC\parallel BC$. Also,
in the degenerate case, if $O$, $B$ and $C$ are collinear, the hypothesis (\ref{equid(O,B,C)}) holds
in general, that is, on one hand we assume more in the hypotheses and prove less in the thesis. These
are some issues of the method and they cannot be fully resolved by using the means we apply.

Now, by multiplying the left hand sides of (\ref{coll(A,O,B)}),
(\ref{equid(O,A,C)}), (\ref{equid(O,B,C)}) and (\ref{perp(A,C,B,C)}), we obtain
\begin{align}\label{product}
\frac{A-O}{O-B}\cdot
\frac{\frac{A-C}{A-O}}{\frac{C-O}{C-A}}\cdot
\frac{\frac{B-O}{B-C}}{\frac{C-B}{C-O}}\cdot
\left(\frac{C-B}{C-A}\right)^2,
\end{align}
which is, after simplifying, $-1$. By assuming that the product of the left hand sides of
(\ref{coll(A,O,B)}), (\ref{equid(O,A,C)}) and (\ref{equid(O,B,C)}) is real,
the remaining left hand side of (\ref{perp(A,C,B,C)}) must also be real to obtain $-1$,
another real number. This proves the statement (manually).

To avoid manual computations and go fully automated, we assign $r_1$, $r_2$ and $r_3$ to
the first three left hand sides (that is, to the hypotheses) and $r$ to the last one (that is, to the thesis).
By considering the elimination ideal $I$ that corresponds to the expressions
$\frac{A-O}{O-B}-r_1$,
$\frac{\frac{A-C}{A-O}}{\frac{C-O}{C-A}}-r_2$,
$\frac{\frac{B-O}{B-C}}{\frac{C-B}{C-O}}-r_3$ and
$\left(\frac{C-B}{C-A}\right)^2-r$, after eliminating the variables $A$, $B$, $C$ and $O$,
we obtain the elimination ideal
\begin{dmath*}
I=\langle r_{2}^{2} r_{3}^{3} r^{2}+2 r_{2}^{2} r_{3}^{2} r-r_{2}^{2} r_{3} r-4 r_{2} r_{3}^{2} r+r_{1}^{2} r_{3}-2 r_{1} r_{2} r_{3}+r_{2}^{2} r_{3}-r_{1} r_{2} r+2 r_{2} r_{3} r+r_{1} r_{2}+4 r_{1} r_{3}-4 r_{2} r_{3}-2 r_{1}+6 r_{3}-4,r_{1}^{2} r_{2}^{2} r^{2}+r_{2}^{2} r_{3}^{2} r^{2}-r_{1}^{2} r_{2}^{2} r+2 r_{1}^{2} r_{2} r+2 r_{2}^{2} r_{3} r+4 r_{1} r_{2} r-r_{2}^{2} r-4 r_{2} r_{3} r+r_{1}^{2}-2 r_{1} r_{2}+r_{2}^{2}+2 r_{2} r+4 r_{1}-4 r_{2}+6,r_{1}^{3} r_{3}-2 r_{1}^{2} r_{2} r_{3}+r_{1} r_{2}^{2} r_{3}-r_{1}^{2} r_{2} r-r_{2} r_{3}^{2} r+r_{1}^{2} r_{2}+4 r_{1}^{2} r_{3}-4 r_{1} r_{2} r_{3}-2 r_{1}^{2}+6 r_{1} r_{3}-2 r_{2} r_{3}-4 r_{1}+r_{2}+4 r_{3}-2,r_{1} r_{2} r_{3} r+1\rangle.
\end{dmath*}
We can identify that $r_1r_2r_3r+1\in I$, that is, $r$ can be expressed as $$r=\frac{-1}{r_1r_2r_3}.$$
If $r_1,r_2,r_3\in\mathbb{R}$, $r$ is also real. We need to double-check, however, that
any of $r_1$, $r_2$ and $r_3$ is different from zero, otherwise $r$ cannot be expressed.
An insertion of the polynomial $r_1r_2r_3$ in the input of the elimination above, we get $I'=\langle1\rangle$
which means that a division by zero cannot occur. This proves the statement mechanically. Clearly,
the obtained result allows the computer to automatically set up the \textit{complex number identity}
$r_1r_2r_3r=-1$ by expressing the constant $-1$ with the rest of the formula, and after substitution this is
the same as
\begin{align} \label{identity-example1}
\frac{A-O}{O-B}\cdot
\frac{\frac{A-C}{A-O}}{\frac{C-O}{C-A}}\cdot
\frac{\frac{B-O}{B-C}}{\frac{C-B}{C-O}}\cdot
\left(\frac{C-B}{C-A}\right)^2=-1.
\end{align}

We need to explain how $I$ was exactly computed. In the theory of algebraic geometry, an
ideal is a ring, therefore, divisions are not allowed. Thus, each division $a_i/b_i$ introduces
a denominator $b_i$. After clearing all denominators (which is a straightforward process in computer algebra)
in each input expression,
an extra slack variable $u$ will be finally introduced, and
then, if there appear $m$ divisions, $b_1\cdot b_2\cdots b_m\cdot u-1$ will be added as an extra polynomial
to the input expressions for the elimination ideal $I$. (This process disallows the denominators to be 0.
This idea is often called ``Rabinowitsch's trick'', referring to the author of \cite{Rabinowitsch_1929}.)
The elimination process must remove $u$ as well. (We provide a detailed computation in Section \ref{maple},
for a simpler example.)

In the rest of the paper we present the process in general: Section \ref{sec:decl-and-realrel} prepares
the main idea, Section \ref{sec:algo} explains our algorithm in details, Section \ref{sec:progs}
demonstrates the computation process in three major computer algebra systems.
Section \ref{sec:gd} presents the implemented algorithm in GeoGebra Discovery.
Section \ref{sec:future} refers to possible future work.

\section{Declarative and real-relational properties}
\label{sec:decl-and-realrel}

We consider a planar construction with points $P_1$, $P_2$, $\ldots$, $P_n$ in $\mathbb{C}$
that is set up by a number of construction steps.

The $i^{\textrm{th}}$ construction step is called \textit{declarative} if there is a polynomial expression $e(P_1,P_2,\ldots,P_{i-1})$
over $\mathbb{Q}$ that defines $P_i$. For example, an alternative construction for Example 1 is the following:
$P_1:=A$ ($A$ is arbitrarily chosen), $P_2:=B$ ($B$ is arbitrarily chosen), $P_3:=(A+B)/2$ (which is $O$), and let $P_4$
(namely, $C$) be on a circle where $AB$ is a diameter (Example 2). Here step 3 is declarative.

The $i^{\textrm{th}}$ construction step is called \textit{real-relational} if there are rational expressions
$e(P_1,P_2,\ldots,P_i)$ over $\mathbb{Q}$ where all of them are real if and only if the property is true.
For example, step 4 can be described with the simple expression $\frac{\frac{P_1-P_4}{P_4-P_3}}{\frac{P_3-P_1}{P_1-P_4}}\in\mathbb{R}$,
because it relies on the angle equality $\angle P_1P_4P_3=\angle P_3P_1P_4$ (which implies $P_1P_3=P_4P_3$,
that is, $AO=CO$), but in Example 1 the same property required two real-relational expressions
(because $OA=OC$ and $OB=OC$ required one complex number identity, respectively). That is, step 4
is real-relational.

Usual construction steps in planar geometry problems include declaring a \textit{midpoint}, the fourth point of a \textit{parallelogram},
the \textit{barycenter} of a triangle (or another polygon); other points that help fulfill
\textit{parallelism}, \textit{collinearity}, \textit{perpendicularity},
\textit{distance equality}, \textit{angle equality},
\textit{concyclicity} or \textit{addition of angles}, can be described by real relations.
See \cite{PengZhangChenLiu_2023} for details on some well-working rational formulas. Here we note that
perpendicularity needs to be handled
together with collinearity. Also, to describe circles with a center $O$ and two circumpoints $P$ and $Q$, we usually set up
a real-relational property that explains angle equality in the isosceles triangle $OPQ$, but in a degenerate
case when $O$, $P$ and $Q$ are collinear, the isosceles property is no longer required. These are issues
that warns us to use the method carefully.

\section{An algorithm}
\label{sec:algo}

\subsection{Computing an elimination ideal and expressing the thesis from the hypotheses}

Having a set of polynomials $p_1,p_2,\ldots$ over complex variables $P_1,P_2,\ldots$ and $r,r_1,r_2,\ldots$,
we can consider the ideal $\langle p_1,p_2\ldots\rangle$ which contains all combinations
$\sum p_i\cdot z_i$ where $z_i\in \mathbb{Q}[P_1,P_2,\ldots,r,r_1,r_2,\ldots]$. This means that
``all polynomial consequences'' of the polynomial equations $p_1=0$, $p_2=0$, $\ldots$ are considered,
by taking the left-hand-sides of the equations having zero on their right-hand-side. Now, if we
restrict the obtained set by considering only those combinations that contain only the variables
$r,r_1,r_2,\ldots$, we can find all possible polynomial equations that hold among these restricted
set of variables.

Such a restriction can be given by the computation
$$\langle p_1,p_2,\ldots\rangle\cap\mathbb{Q}[r,r_1,r_2,\ldots]$$
and the result can be expressed as an ideal $I$ that is generated with a finite number of polynomials
from $\mathbb{Q}[r,r_1,r_2,\ldots]$. This is ensured by \textit{elimination theory}
(see \cite{Cox_2007}, Chapter 3), and it can be proven that there are finite algorithms that compute
the \textit{elimination ideal} $I$ by a well-chosen term ordering when a Gröbner basis
of the input polynomials is calculated. In the last decades, the existing algorithms have been
improved remarkably, so we can rely on fast computation of the elimination ideals, even if the
worst case still requires double exponential basic operations on polynomials, in the number of variables
\cite{MayrMeyer82}.

$I$ can be obtained in a form where $r$ is expressed so that the degree of $r$ is minimal,
by running the algorithm with the most appropriate term ordering. Therefore, by selecting
the minimal degree polynomial of $r$ from $I$, we can express $r$ in a simple way by using only the variables $r_1,r_2,\ldots$
Assuming that $r$ is linear in this form, we can express $r$ by a rational expression.
If we can exclude the case when the denominator is zero,
then we can argue that $r$ can always be expressed (except for a couple of counterexamples) with $r_1,r_2,\ldots$
If such an expression contains only real values, then $r$ itself must be a real number.

\subsection{A basic algorithm}

Now we can define an algorithm to decide the truth of a planar geometry statement (the thesis $t$) which needs to
be real-relational in our approach:

\begin{enumerate}
\item Collect all points that are created during the construction in a declarative way, and substitute
each point $P_i$ by its declaration $e(P_1,P_2,\ldots,P_{i-1})$. (From the computer algebra point of view,
this means a simple declaration of point $P_i$.)
\item Collect all points that are created during the construction in a real-relational way, and for each relation
create a rational expression $e(P_1,P_2,\ldots,P_{i})-r_i$.
\item For the thesis $t$, create a rational expression $e(P_1,P_2,\ldots,P_{n})-r$ where $n$ is the number of
construction steps.
\item Clear all denominators $b_1$, $b_2$, $\ldots$, $b_m$, from step 2 and 3
to obtain polynomials $p_1$, $p_2$, $\ldots$, $p_s$ and
consider a slack variable $u$ to create the polynomial $p_{s+1}=b_1\cdot b_2\cdots b_m\cdot u-1$.
(This step is automatically supported in Mathematica/WolframScript, and also in the newest versions of Giac.)
\item Compute the elimination ideal
\begin{align}\label{elimideal}
I=\langle p_1,p_2,\ldots,p_s,p_{s+1}\rangle \cap \mathbb{Q}[r,r_1,r_2,\ldots].
\end{align}
Technically, this means the elimination of those variables that do not correspond to a declarative step,
and also $u$ must be eliminated.
\item If no polynomials of $I$ contain $r$, then the statement $t$ cannot be proven.
(This usually means that the statement
is false, or, the thesis is not a clear conclusion of the hypotheses. But, at this point, we do not
have a proof of falsity.) Stop.
\item Let $d$ be the minimal degree of $r$ among the polynomials in $I$.
\item If $d=1$, then we express $r$ in a linear expression and we can obtain a rational form for $r$.
\begin{enumerate}
\item If the rational form is polynomial, then the thesis is true. Stop.
\item Otherwise let the denominator in the rational form of $r$ be denoted by $D(r_1,r_2,\ldots)$.
\item Compute
\begin{align}\label{elimideal}
I'=\langle p_1,p_2,\ldots,p_s,p_{s+1},D\rangle \cap \mathbb{Q}[r,r_1,r_2,\ldots],
\end{align}
that is, repeat the elimination with inserting $D$ in the input.
\item If $I'=\langle1\rangle$, then the statement $t$ is true (because if $D=0$, we get a contradiction). Stop.
\item If no polynomials of $I'$ contain $r$, then the statement $t$ cannot be proven (because if $D=0$, the thesis
is not a clear conclusion of the hypotheses). Stop.
\item Let $d'$ be the minimal degree of $r$ among the polynomials in $I'$.
\item If $d'=1$, then we express $r$ in a second linear expression and we can obtain a second rational form for $r$.
\begin{enumerate}
\item If the rational form is polynomial, then the thesis is true (because if $D=0$, we still have
a rational form for $r$ in general, except for a couple of counterexamples). Stop.
\end{enumerate}
\end{enumerate}
\item In any other cases, the algorithm is not conclusive.
\end{enumerate}

To explain the most important part of the algorithm, we need to show that in step 8 we indeed have a proof.
Because $r$ appears in a linear expression
like $r\cdot v+w(=0)$ (where $v$ and $w$ are polynomials of $r_1$, $r_2$, $\ldots$),
it can be expressed by the formula $r=-w/v$. Since $v,w\in\mathbb{R}$, $r$ is also real. That is,
assuming that all the declarative and real-relational properties hold, the thesis must also hold.

The algorithm is finite, because computing the elimination ideal is finite. However, it may take a long time,
depending on the number of variables and on the input in general.

\subsection{A practical improvement}

It seems logical assuming $A:=0$, $B:=1$ (or, in some symmetric cases, $A:=-1$, $B:=1$)
to save two variables, since elimination may be a slow procedure in general. (In its slowest
case it is double exponential in the number of variables.) This indeed speeds up computation
in certain cases.

\subsection{Automated creation of readable proofs}

A major benefit of the complex number identity method that geometric proofs can be explained
in a readable way. By using the above mentioned algorithm, it is quite straightforward to
create such a proof automatically. The computer does not have to do anything but following the
algorithm, expressing $r$, and arguing why it must be a real number if $r_1,r_2,\ldots\in\mathbb{R}$.

\section{Computer programs}
\label{sec:progs}

In this section we provide working computer programs that produce an automatic proof of the examples
mentioned above.

\subsection{Implementation in Giac}

Giac is a computer algebra developed by Bernard Parisse
(Institut Fourier, Université de Grenoble, France).
It is free of charge and part of the Debian Linux
system (package \texttt{xcas}) and its derivatives.
It is also available online at \url{https://www-fourier.univ-grenoble-alpes.fr/~parisse/giac/xcas.html}.

To run the following programs without an issue, one needs a recent version
of Giac that handles divisions and extends the input automatically as described above.

\subsubsection*{Example 1}

\begin{lstlisting}[language=giac]
1>> I:=eliminate([(A-O)/(O-B)-r1,(A-C)/(A-O)/((C-O)/(C-A))-r2,(B-O)/(B-C)/((C-B)/(C-O))-r3,((C-B)/(C-A))^2-r],[A,B,C,O])
[r2^2*r3^3*r^2+2*r2^2*r3^2*r-r2^2*r3*r-4*r2*r3^2*r+r1^2*r3-2*r1*r2*r3+r2^2*r3-r1*r2*r+2*r2*r3*r+r1*r2+4*r1*r3-4*r2*r3-2*r1+6*r3-4,r1^2*r2^2*r^2+r2^2*r3^2*r^2-r1^2*r2^2*r+2*r1^2*r2*r+2*r2^2*r3*r+4*r1*r2*r-r2^2*r-4*r2*r3*r+r1^2-2*r1*r2+r2^2+2*r2*r+4*r1-4*r2+6,r1^3*r3-2*r1^2*r2*r3+r1*r2^2*r3-r1^2*r2*r-r2*r3^2*r+r1^2*r2+4*r1^2*r3-4*r1*r2*r3-2*r1^2+6*r1*r3-2*r2*r3-4*r1+r2+4*r3-2,r1*r2*r3*r+1]
2>> degree(I[3],r)
2<< 1
3>> solve(I[3],r)
3<< list[-1/(r1*r2*r3)]
4>> I:=eliminate([(A-O)/(O-B)-r1,(A-C)/(A-O)/((C-O)/(C-A))-r2,(B-O)/(B-C)/((C-B)/(C-O))-r3,((C-B)/(C-A))^2-r,r1*r2*r3],[A,B,C,O])
4<< [1]
\end{lstlisting}

\subsubsection*{Example 2}

\begin{lstlisting}[language=giac]
1>> equid(P,Q,R):=(Q-R)/(P-R)/((P-Q)/(R-Q));
2>> perp(P,Q,R,S):=((P-Q)/(R-S))^2;
3>> O:=(A+B)/2;
4>> I:=eliminate([equid(O,A,C)-r1,perp(A,C,C,B)-r],[A,B,C])
4<< [r1*r-r1-4*r]
5>> degree(I[0],r)
5<< 1
6>> solve(I[0],r)
6<< list[r1/(r1-4)]
7>> I:=eliminate([equid(O,A,C)-r1,perp(A,C,C,B)-r,r1-4],[A,B,C])
7<< [1]
\end{lstlisting}

\subsection{Mathematica}

Here we provide only one program for reference. Accordingly,
the other example written in Giac can be translated to Mathematica.

\subsubsection*{Example 2}

\begin{lstlisting}[language=mathematica]
In[1]:= equid[P_,Q_,R_]:=(Q-R)/(P-R)/((P-Q)/(R-Q))
In[2]:= perp[P_,Q_,R_,S_]:=((P-Q)/(R-S))^2
In[3]:= O1:=(A+B)/2
In[4]:= I1=GroebnerBasis[{equid[O1,A,C]-r1,perp[A,C,C,B]-r},{},{A,B,C},MonomialOrder->EliminationOrder]//InputForm
Out[4]//InputForm= {-4*r - r1 + r*r1}
In[5]:= Exponent[I1[[1]],r]
Out[5]= {1}
In[6]:= Solve[I1[[1]]==0,r]
                 r1
Out[6]= {{r -> -------}}
               -4 + r1
In[7]:= I2=GroebnerBasis[{equid[O1,A,C]-r1,perp[A,C,C,B]-r,-4+r1},{},{A,B,C},MonomialOrder->EliminationOrder]//InputForm
Out[7]//InputForm= {1}
\end{lstlisting}

\subsection{Maple}
\label{maple}

Maple does not allow in the input of the command \texttt{EliminationIdeal} that the
user provides expressions that contain divisions. Therefore, we need to manually
extend the input accordingly.

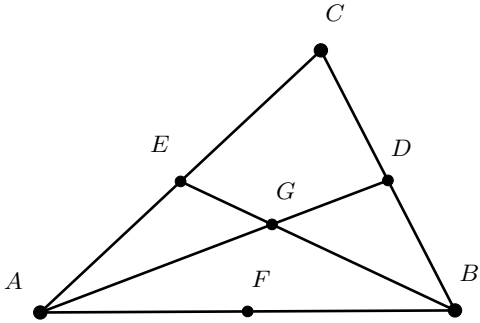
\begin{figure}[ht]
\definecolor{uuuuuu}{rgb}{0,0,0}
\definecolor{ududff}{rgb}{0,0,0}
\begin{tikzpicture}[line cap=round,line join=round,>=triangle 45,x=5.0cm,y=5.0cm]
\clip(0.8084213835580526,0.6365486255803741) rectangle (2.20384974287229,1.6790476333170805);
\draw [line width=1.pt] (0.976741535848875,0.8102984602031584)-- (2.0681076845732393,0.8157281425351205);
\draw [line width=1.pt] (2.0681076845732393,0.8157281425351205)-- (1.7151783329957087,1.5052977986942961);
\draw [line width=1.pt] (1.7151783329957087,1.5052977986942961)-- (0.976741535848875,0.8102984602031584);
\draw [line width=1.pt] (0.976741535848875,0.8102984602031584)-- (1.891643008784474,1.1605129706147084);
\draw [line width=1.pt] (2.0681076845732393,0.8157281425351205)-- (1.345959934422292,1.1577981294487274);
\begin{scriptsize}
\draw [fill=ududff] (0.976741535848875,0.8102984602031584) circle (2.5pt);
\draw[color=ududff] (0.9061556655333688,0.8944585363485696) node {$A$};
\draw [fill=ududff] (2.0681076845732393,0.8157281425351205) circle (2.5pt);
\draw[color=ududff] (2.1061154608969734,0.9161772656764178) node {$B$};
\draw [fill=ududff] (1.7151783329957087,1.5052977986942961) circle (2.5pt);
\draw[color=ududff] (1.7531861093194427,1.6057469218355933) node {$C$};
\draw [fill=uuuuuu] (1.891643008784474,1.1605129706147084) circle (2.0pt);
\draw[color=uuuuuu] (1.9269359439422271,1.2473878879261004) node {$D$};
\draw [fill=uuuuuu] (1.345959934422292,1.1577981294487274) circle (2.0pt);
\draw[color=uuuuuu] (1.2916631111026717,1.2582472525900243) node {$E$};
\draw [fill=uuuuuu] (1.5224246102110572,0.8130133013691394) circle (2.0pt);
\draw[color=uuuuuu] (1.5577175453688101,0.8998882186805317) node {$F$};
\draw [fill=uuuuuu] (1.586675851139275,1.0437748004775254) circle (2.0pt);
\draw[color=uuuuuu] (1.6228737333523544,1.1333645589548982) node {$G$};
\end{scriptsize}
\end{tikzpicture}
\caption{Example 3: Medians of a triangle are concurrent}
\label{fig:7}
\end{figure}

(Example 3.) We prove that the medians of a triangle are concurrent (Fig.~\ref{fig:7}).
We consider $\triangle ABC$
with $P_1:=A$, $P_2:=B$, $P_3:=C$ and the declarations $P_4:=D$, $P_5:=E$, $P_6:=F$
where $D=(B+C)/2$, $E=(A+C)/2$, $F=(A+B)/2$. Let $P_7:=G$ such that $G$ is
the intersection of $AD$ and $BE$, that is, $A$, $G$ and $D$ are collinear,
and the same is true for $B$, $G$ and $E$. Therefore, the real-relational properties
$(B-G)/(B-E)\in\mathbb{R}$ and $(D-G)/(D-A)\in\mathbb{R}$ hold.
To show: $(C-G)/(C-F)\in\mathbb{R}$.

Now, to create an acceptable input for Maple, we manually rewrite the real-relational
expressions and obtain the input ideal
$$
\langle (B-E)r_1+G-B,(D-A)r_2+G-D,(C-F)r+G-C,\\
u\cdot (B-E)(D-A)(C-F)-1\rangle,
$$
which, after intersecting with $\mathbb{Q}[r,r_1,r_2]$, gives the output
$I=\langle-3 r r_1 + 3 r r_2 + 3 r_1 r_2 + r + r_1 - 4 r_2\rangle$. This is linear in $r$,
therefore, the algorithm can continue with the next steps.

The complete Maple code of these first steps looks as follows:
\begin{lstlisting}[language=maple]
with(PolynomialIdeals):
D1:=(B+C)/2:
E:=(A+C)/2:
F:=(A+B)/2:
EliminationIdeal(<(B-E)*r1+G-B,(D1-A)*r2+G-D1,(C-F)*r+G-C,u*(B-E)*(D1-A)*(C-F)-1>,{r1,r2,r});
\end{lstlisting}

\section{A prototype in GeoGebra Discovery}
\label{sec:gd}

We integrated the explained algorithm in the free dynamic geometry system,
GeoGebra Discovery. Since Giac is embedded into that program \cite{GiacGG-RICAM2013}, we use Giac
to compute the elimination ideal and the other steps of the algorithm.
When starting the desktop version of the program with the command line
argument \texttt{--prover=engine:cni}, the new CNI (``complex number identity'') method will be used when
the command \textbf{Prove} is issued by the user.

The implementation provides full integration in GeoGebra Discovery's automated reasoning subsystem.
This means that all other commands that based on the \textbf{Prove} command,
including \textbf{Discover} \cite{rmec,EPTCS354.1} or \textbf{ShowProof} \cite{showproof-ISSAC2024}, will directly use the new method.

\subsection{Example 4: The midpoint parallelogram}

As an example input we consider an arbitrary quadrilateral $ABCD$ in the plane.
When constructing the midpoints of each side, we obtain points $E$, $F$, $G$ and $H$.
It is well-known that the quadrilateral $EFGH$ always form a parallelogram:
this fact is also called Varignon's theorem, originally published in 1731 \cite{Varignon_1731}.

\begin{figure}
\includegraphics[width=0.7\textwidth]{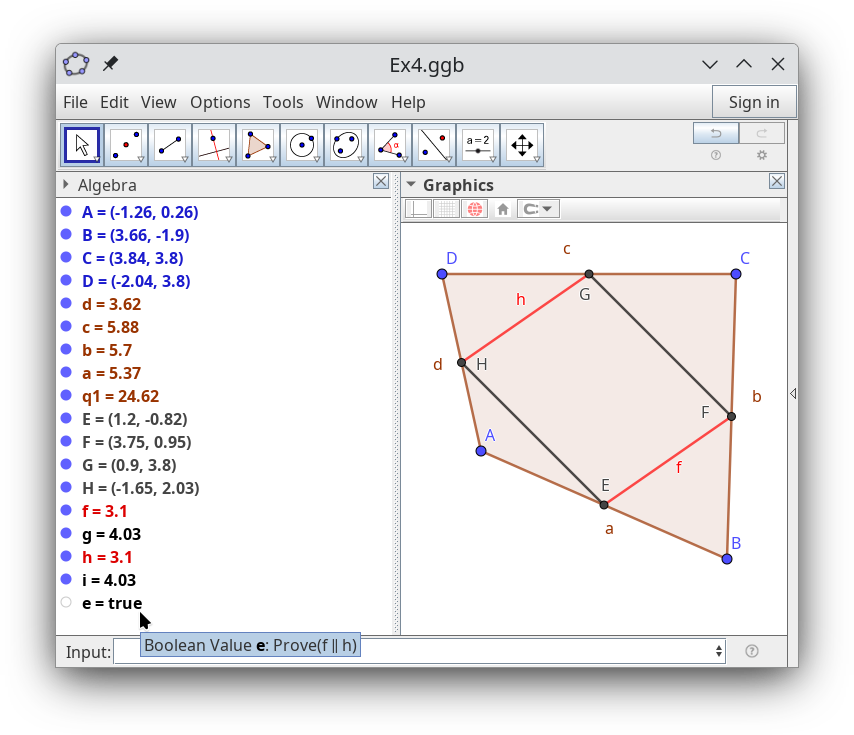}
\caption{Example 4: The midpoint parallelogram in GeoGebra Discovery}
\label{fig:4}
\end{figure}

Here we show how this theorem can be studied in GeoGebra Discovery (Fig.~\ref{fig:4}). By entering
\texttt{Prove($f \parallel h$)} in the input field, an almost immediate result is shown (it takes
7 ms on an AMD Ryzen 5 5600H processor, under Ubuntu Linux 24.04). When issuing the command
\texttt{ShowProof($f \parallel h$)}, the program gives the detailed explanation as follows
(here we show the \LaTeX\ export of the CAS View):

\begin{tcolorbox}[
    enhanced, 
    breakable,
    colframe=black!50,
    colback=black!10,
    colbacktitle=black!40,
    colupper=black,
    title={The output of the CAS View for the input \texttt{ShowProof($f \parallel h$)}
    for the midpoint parallelogram theorem}
    ]
\begin{enumerate}
\item
\definecolor{mycolor0}{rgb}{0.0, 0.0, 0.99609375}
\textcolor{mycolor0}{Let A, B, C, D be arbitrary points.}
\item
\definecolor{mycolor1}{rgb}{0.0, 0.0, 0.99609375}
\textcolor{mycolor1}{Let q1 be the polygon A, B, C, D.}
\item
\definecolor{mycolor2}{rgb}{0.0, 0.0, 0.99609375}
\textcolor{mycolor2}{Let a be the segment A, B.}
\item
\definecolor{mycolor3}{rgb}{0.0, 0.0, 0.99609375}
\textcolor{mycolor3}{Let b be the segment B, C.}
\item
\definecolor{mycolor4}{rgb}{0.0, 0.0, 0.99609375}
\textcolor{mycolor4}{Let c be the segment C, D.}
\item
\definecolor{mycolor5}{rgb}{0.0, 0.0, 0.99609375}
\textcolor{mycolor5}{Let d be the segment D, A.}
\item
\definecolor{mycolor6}{rgb}{0.0, 0.0, 0.99609375}
\textcolor{mycolor6}{Let E be the midpoint of a.}
\item
\definecolor{mycolor7}{rgb}{0.0, 0.0, 0.99609375}
\textcolor{mycolor7}{Let F be the midpoint of b.}
\item
\definecolor{mycolor8}{rgb}{0.0, 0.0, 0.99609375}
\textcolor{mycolor8}{Let G be the midpoint of c.}
\item
\definecolor{mycolor9}{rgb}{0.0, 0.0, 0.99609375}
\textcolor{mycolor9}{Let H be the midpoint of d.}
\item
\definecolor{mycolor10}{rgb}{0.0, 0.0, 0.99609375}
\textcolor{mycolor10}{Let f be the segment E, F.}
\item
\definecolor{mycolor11}{rgb}{0.0, 0.0, 0.99609375}
\textcolor{mycolor11}{Let h be the segment G, H.}
\item
\definecolor{mycolor12}{rgb}{0.0, 0.0, 0.99609375}
\textcolor{mycolor12}{Prove that f  $\parallel$  h.}
\item
\definecolor{mycolor13}{rgb}{0.99609375, 0.0, 0.0}
\textcolor{mycolor13}{The statement is true under some non-degeneracy conditions (see below).}
\item
\definecolor{mycolor14}{rgb}{0.0, 0.0, 0.0}
\textcolor{mycolor14}{The hypotheses:}
\item
\definecolor{mycolor15}{rgb}{0.0, 0.0, 0.0}
\textbf{\textcolor{mycolor15}{E:=(A+B)/2}}
\item
\definecolor{mycolor16}{rgb}{0.0, 0.0, 0.0}
\textbf{\textcolor{mycolor16}{F:=(B+C)/2}}
\item
\definecolor{mycolor17}{rgb}{0.0, 0.0, 0.0}
\textbf{\textcolor{mycolor17}{G:=(C+D)/2}}
\item
\definecolor{mycolor18}{rgb}{0.0, 0.0, 0.0}
\textbf{\textcolor{mycolor18}{H:=(D+A)/2}}
\item
\definecolor{mycolor19}{rgb}{0.99609375, 0.49609375, 0.0}
\textcolor{mycolor19}{Without loss of generality, some coordinates can be fixed:}
\item
\definecolor{mycolor20}{rgb}{0.0, 0.0, 0.0}
\textbf{\textcolor{mycolor20}{A:=0}}
\item
\definecolor{mycolor21}{rgb}{0.0, 0.0, 0.0}
\textbf{\textcolor{mycolor21}{B:=1}}
\item
\definecolor{mycolor22}{rgb}{0.0, 0.0, 0.0}
\textcolor{mycolor22}{The thesis:}
\item
\definecolor{mycolor23}{rgb}{0.0, 0.0, 0.0}
\textcolor{mycolor23}{f  $\parallel$  h}
\item
\definecolor{mycolor24}{rgb}{0.0, 0.0, 0.0}
\textbf{\textcolor{mycolor24}{(E-F)/(G-H)=r}}
\item
\definecolor{mycolor25}{rgb}{0.0, 0.0, 0.0}
\textcolor{mycolor25}{We eliminate all variables that correspond to complex points.}
\item
\definecolor{mycolor26}{rgb}{0.0, 0.0, 0.0}
\textcolor{mycolor26}{The thesis (r) can be expressed as a rational expression of the hypotheses, because r is linear in an obtained polynomial equation:}
\item
\definecolor{mycolor27}{rgb}{0.0, 0.0, 0.0}
\textcolor{mycolor27}{-r-1=0}
\item
\definecolor{mycolor28}{rgb}{0.0, 0.0, 0.0}
\textcolor{mycolor28}{The thesis can be expressed as a polynomial expression of the hypotheses.}
\item
\definecolor{mycolor29}{rgb}{0.99609375, 0.0, 0.0}
\textbf{\textcolor{mycolor29}{Since all hypotheses are real expressions, the thesis must also be real.}}
\end{enumerate}
\end{tcolorbox}

Analogously, a similar check can be done for checking parallelism of $FG$ and $HE$.

\subsection{Example 5: Angle bisectors of a triangle are concurrent}

\begin{figure}
\includegraphics[width=0.7\textwidth]{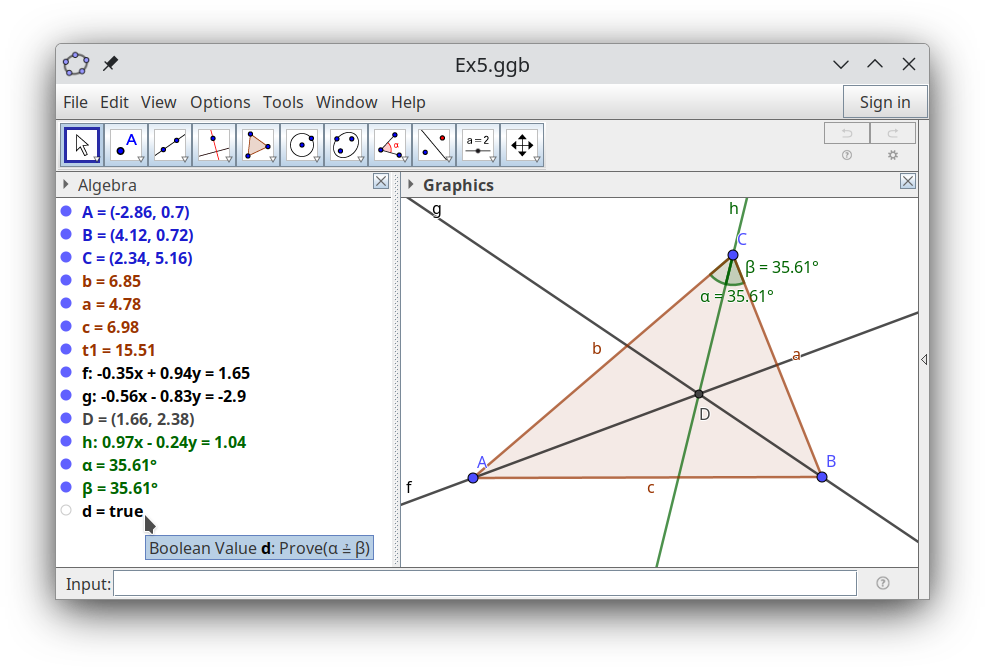}
\caption{Example 4: Angle bisectors of a triangle are concurrent, constructed in GeoGebra Discovery}
\label{fig:5}
\end{figure}

As a final example, we prove that the angle bisectors in a triangle are always concurrent.
The construction in GeoGebra Discovery can be seen in Fig.~\ref{fig:5}: Point $D$ is defined
as the intersection of angle bisectors $f$ (through $A$) and $g$ (through $B$), and
angles $\alpha$ and $\beta$ are given as $\angle ACD$ and $\angle DCB$, respectively.

By entering
\texttt{Prove($\alpha == \beta$)} in the input field, the result is shown in 141 ms on the same computer.
When issuing the command
\texttt{ShowProof($f \parallel h$)}, the program gives the following explanation:

\begin{tcolorbox}[
    enhanced, 
    breakable,
    colframe=black!50,
    colback=black!10,
    colbacktitle=black!40,
    colupper=black,
    title={The output of the CAS View for the input \texttt{ShowProof($\alpha == \beta$)}
    for the concurrency of angle bisectors}
    ]
\begin{enumerate}
\item
\definecolor{mycolor0}{rgb}{0.0, 0.0, 0.99609375}
\textcolor{mycolor0}{Let A, B, C be arbitrary points.}
\item
\definecolor{mycolor1}{rgb}{0.0, 0.0, 0.99609375}
\textcolor{mycolor1}{Let f be the angle bisector of B, A, C.}
\item
\definecolor{mycolor2}{rgb}{0.0, 0.0, 0.99609375}
\textcolor{mycolor2}{Let g be the angle bisector of A, B, C.}
\item
\definecolor{mycolor3}{rgb}{0.0, 0.0, 0.99609375}
\textcolor{mycolor3}{Let D be the intersection of f and g.}
\item
\definecolor{mycolor4}{rgb}{0.0, 0.0, 0.99609375}
\textcolor{mycolor4}{Denote the expression Angle between A, C, D by $\alpha$.}
\item
\definecolor{mycolor5}{rgb}{0.0, 0.0, 0.99609375}
\textcolor{mycolor5}{Denote the expression Angle between D, C, B by $\beta$.}
\item
\definecolor{mycolor6}{rgb}{0.0, 0.0, 0.99609375}
\textcolor{mycolor6}{Prove that $\alpha$  $\stackrel{?}{=}$  $\beta$.}
\item
\definecolor{mycolor7}{rgb}{0.99609375, 0.0, 0.0}
\textcolor{mycolor7}{The statement is true under some non-degeneracy conditions (see below).}
\item
\definecolor{mycolor8}{rgb}{0.0, 0.0, 0.0}
\textcolor{mycolor8}{The hypotheses:}
\item
\definecolor{mycolor9}{rgb}{0.0, 0.0, 0.0}
\textcolor{mycolor9}{Considering definition D = Intersect(f, g):}
\item
\definecolor{mycolor10}{rgb}{0.0, 0.0, 0.0}
\textbf{\textcolor{mycolor10}{(B-A)/(B-D)/((B-D)/(B-C))=r1 $\in$ $\mathbb{R}$}}
\item
\definecolor{mycolor11}{rgb}{0.0, 0.0, 0.0}
\textbf{\textcolor{mycolor11}{(A-B)/(A-D)/((A-D)/(A-C))=r2 $\in$ $\mathbb{R}$}}
\item
\definecolor{mycolor12}{rgb}{0.99609375, 0.49609375, 0.0}
\textcolor{mycolor12}{Without loss of generality, some coordinates can be fixed:}
\item
\definecolor{mycolor13}{rgb}{0.0, 0.0, 0.0}
\textbf{\textcolor{mycolor13}{A:=0}}
\item
\definecolor{mycolor14}{rgb}{0.0, 0.0, 0.0}
\textbf{\textcolor{mycolor14}{B:=1}}
\item
\definecolor{mycolor15}{rgb}{0.0, 0.0, 0.0}
\textcolor{mycolor15}{The thesis:}
\item
\definecolor{mycolor16}{rgb}{0.0, 0.0, 0.0}
\textcolor{mycolor16}{$\alpha$  $\stackrel{?}{=}$  $\beta$}
\item
\definecolor{mycolor17}{rgb}{0.0, 0.0, 0.0}
\textbf{\textcolor{mycolor17}{(C-A)/(C-D)/((C-D)/(C-B))=r}}
\item
\definecolor{mycolor18}{rgb}{0.0, 0.0, 0.0}
\textcolor{mycolor18}{We eliminate all variables that correspond to complex points.}
\item
\definecolor{mycolor19}{rgb}{0.0, 0.0, 0.0}
\textcolor{mycolor19}{The thesis (r) can be expressed as a rational expression of the hypotheses, because r is linear in an obtained polynomial equation:}
\item
\definecolor{mycolor20}{rgb}{0.0, 0.0, 0.0}
\textcolor{mycolor20}{r1*r2*r-r1*r2-r1*r-r2*r=0}
\item
\definecolor{mycolor21}{rgb}{0.0, 0.0, 0.0}
\textcolor{mycolor21}{Expressing the thesis requires a division by r1*r2-r1-r2.}
\item
\definecolor{mycolor22}{rgb}{0.0, 0.0, 0.0}
\textcolor{mycolor22}{Let us assume that that divisor is 0 and restart the elimination.}
\item
\definecolor{mycolor23}{rgb}{0.0, 0.0, 0.0}
\textcolor{mycolor23}{The elimination verifies that that divisor cannot be zero.}
\item
\definecolor{mycolor24}{rgb}{0.99609375, 0.0, 0.0}
\textbf{\textcolor{mycolor24}{Since all hypotheses are real expressions, the thesis must also be real.}}
\end{enumerate}
\end{tcolorbox}

\subsection{Implementation details}

The prototype of the CNI method has been implemented in a single Java file, in the class ProverCNIMethod in the Java package
\texttt{org.geogebra.common.prover} in less than 900 lines of code. Unlike other proving methods that are strewn in various
files in the source code, the CNI method is a compact algorithm which can be considered relatively simple and
well-maintainable. It also contains the explanatory comments that are displayed in the output of the \texttt{ShowProof} command.

The prototype does not identify false statements at the moment. It classifies all statements that cannot be proven
as ``unknown'', to stay on the safe side.

We tested the prototype with various inputs, including 
the prover benchmark database of GeoGebra Discovery. When running the full test
(which consists of 353 test cases at the moment\footnote{See
\url{https://prover-test.risc.jku.at/job/GeoGebra_Discovery-provertest/136/artifact/fork/geogebra/test/scripts/benchmark/prover/html/all.html}
for the benchmarking outputs for the current version, including the outputs for some alternative provers as well.},
some of them are intentionally false statements), we found the following results
(in quotation marks at the end of each item we refer to the classification codes in the benchmark output):
\begin{itemize}
\item 20 test cases worked properly,
\item 49 cases ran into timeout (which was set to 20 seconds), ``t/o'',
\item 248 cases failed because of at least one missing implementation among the construction steps, ``niu'',
\item 15 cases failed because of at least one incomplete implementation among the construction steps, ``nfiu''.
\item In 5 cases the targeted check was to prove that $r=0$ (this is required to prove point equality), but this check failed, ``rn0u''.
\item In 8 cases, the algorithm found that $r$ cannot be expressed in a linear way, ``nlu''.
\item In 3 cases, it seemed useful to try another (a third) elimination to study the situation even more,
but this check was not yet built neither in the prototype, nor in the algorithm, ``d3u''.
\item In 3 cases, the elimination ideal was $\langle0\rangle$, that is, no conclusion could be found, ``e0u''.
Other than that, there was no case when $r$ was not present in the elimination ideal.
\item In 2 cases, however, $r$ was not present in the second elimination ideal, ``e2nru''.
\end{itemize}

When comparing the properly working cases to GeoGebra Discovery's default prover (\textit{Botana's method}
\cite{sws-eaca-paper}, it is
definitely a ``Gröbner basis'' method that uses two coordinates for each introduced point), it is clear
that the CNI method may outperform the default prover in several test cases.
In fact, in the number of variables, each method usually requires
two variables for each introduced point: the CNI method needs one complex number and one real variable to
introduce a real relation, while Botana's method needs two coordinates. The CNI method can, however, save
some variables if certain steps are declarative. Also, it is important to construct the test case in such
a way so that the underlying method does not have to introduce new points or extra relations. It seems that
introducing the non-degeneracy condition (to avoid divisions in the input of elimination process) may lead
to a high degree polynomial in several variables, and this may be a root of infeasibly long computations.
This conjecture should be, however, properly checked as future work.

The latest version of GeoGebra Discovery can be found at \url{https://github.com/kovzol/geogebra-discovery/}.

\section{Discussion and future work}
\label{sec:future}

The method we presented in this paper seems to be useful to solve certain problems in plane geometry, but it
is clearly not suitable to generalize in 3 dimensions, at least not in a trivial way. It needs to be used carefully,
since some relations may refer to a wider set of properties than expected, especially when degenerate cases arise.

The implemented prototype should be further improved to handle a lot more construction variants in GeoGebra. We expect
a substantial progress when a wider set of construction steps will be fully supported in the code. The first impressions
give us hope to continue the work on the automated method as well, by handling such cases that are not yet fully covered by
the theory. This includes the cases when $r$ cannot be expressed in a linear way, but is of a higher degree. These
cases may require a more refined study when expressing $r$ with the real variables $r_1,r_2,\ldots$

One bottleneck of the method is the slow computation of the elimination ideal. We will try to improve
the computer algebra algorithms to speed up the computations in the future. Also, it seems relevant to find
the most appropriate form of algebraization since there are usually more options: for example, collinearity can be
expressed in multiple ways when permuting the input variables.

\section*{Acknowledgments}

Vilmos Molnár-Szabó kindly read the first draft of this paper and gave useful feedback.
We are thankful to Bernard Parisse who improved Giac to make it possible to compute the elimination ideal
in a simple way, without dealing with clearing the denominators manually.

\newcommand{\doi}[1]{{doi:#1}}

\bibliography{kovzol-publications,extra}

\end{document}